\documentclass[letter,aps,reprint]{revtex4-1}

\usepackage{amsmath}
\usepackage{amssymb}
\usepackage{braket}
\usepackage{graphicx}
\usepackage{textgreek}
\usepackage{textcomp}
\usepackage{xcolor}
\usepackage{placeins}
\usepackage{hyperref}
\usepackage[resetlabels]{multibib}
\newcites{M}{methods references}

\newcommand{\ad}{\hat{a}^{\dag}}
\renewcommand{\a}{\hat{a}}
\newcommand{\cd}{\hat{c}^{\dag}}
\renewcommand{\c}{\hat{c}}

\newcommand{\Dc}{{\Delta}_\mathrm{c}}
\newcommand{\Dct}{\tilde{\Delta}_\mathrm{c}}

\newcommand{\kt}{\mathbf{k}_x}
\newcommand{\kc}{\mathbf{k}_\mathrm{c}}

\renewcommand{\r}{\mathbf{r}}

\renewcommand{\k}{\mathbf{k}}

\newcommand{\Da}{\Delta_\mathrm{a}}
\renewcommand{\Re}{\mathbb{R}\mathrm{e}}
\renewcommand{\Im}{\mathbb{I}\mathrm{m}}

\renewcommand{\P}{\hat{\Psi}}
\newcommand{\Pt}{\hat{\Psi}^{\dag}}
\newcommand{\dr}{\mathrm{d}\r\,}
\newcommand{\hatTe}{\hat{\Theta}_q}
\newcommand{\hatTo}{\hat{\Theta}_p}
\newcommand{\Te}{\Theta_q}
\newcommand{\To}{\Theta_p}
\newcommand{\er}{\eta_q}
\newcommand{\ei}{\eta_p}

\newcommand{\ETH}{Institute for Quantum Electronics, Eidgen\"ossische Technische Hochschule Z\"urich, Otto-Stern-Weg 1, 8093 Zurich, Switzerland}

\begin{document}

\title{Self-oscillating  pump in a topological dissipative atom-cavity system}

\author{Davide Dreon}
\author{Alexander Baumg\"artner}
\author{Xiangliang Li}
\author{Simon Hertlein}
\author{Tilman Esslinger}
\email{esslinger@phys.ethz.ch}
\author{Tobias Donner}
\affiliation{\ETH}

\maketitle

\textbf{Pumps are transport mechanisms in which direct currents result from a cyclic evolution of the potential~\cite{Altshuler:1999aa,cohen2003quantum}. As Thouless has shown, the pumping process can have topological origins, when considering the motion of quantum particles in spatially and temporally periodic potentials~\cite{thouless1983quantization}. 
However, the periodic evolution that drives these pumps has always been assumed to be imparted from outside, as was the case in the experimental systems studied so far~\cite{switkes1999adiabatic,aleiner1998adiabatic,blumenthal2007gigahertz,giazotto2011josephson,lu2016geometrical,nakajima2016topological,lohse2016thouless,lohse2018exploring,nakajima2021competition}. Here we report on an emergent mechanism for pumping in a quantum gas coupled to an optical resonator, where we observe a particle current without applying a periodic drive. The pumping potential experienced by the atoms is formed by the self-consistent cavity field interfering with the static laser field driving the atoms. Due to dissipation, the cavity field evolves between its two quadratures~\cite{Dogra2019diss}, each corresponding to a different centrosymmetric crystal configuration~\cite{li2021first}.  This self-oscillation results in a time-periodic potential analogous to that describing the transport of electrons in topological tight-binding models, like the paradigmatic Rice-Mele pump~\cite{rice1982elementary}. In the experiment, we directly follow the evolution by measuring the phase winding of the cavity field with respect to the driving field and observing the atomic motion in-situ. 
The discovered mechanism combines the dynamics of topological and open systems, and features characteristics of continuous dissipative time crystals.}

Models for geometrical pumps in lattices show singularities in their parameter space, which lead to pumping as soon as these are encircled~\cite{xiao2010berry,Altshuler:1999aa}, with the transported charge proportional to the winding number. For filled bands the current is quantized in this number and topologically protected, as recognized by Thouless~\cite{thouless1983quantization}. Such singularities appear in lattices with bipartite unit cells, for example in the Rice-Mele model~\cite{rice1982elementary}. This paradigmatic model describes a tight binding lattice with alternating onsite energies and hopping integrals, where the singularity corresponds to the simultaneous degeneracies of both. If the singularity is periodically encircled by these two parameters, transport can be induced. This is the basis for topological pumps realized in cold atoms, where a one-dimensional optical lattice is overlaid by a second lattice with twice the spatial period~\cite{niu1986quantum,wang2013topological,qian2011quantum,nakajima2016topological,lohse2016thouless}. An external drive then slowly moves the latter relative to the former along the joint lattice axis. 
The pumping follows a cyclic evolution after which the potential landscape resembles itself with the centre of the Wannier functions being shifted by a unit vector. 

\begin{figure}[h]
\centering
\includegraphics[width = \columnwidth]{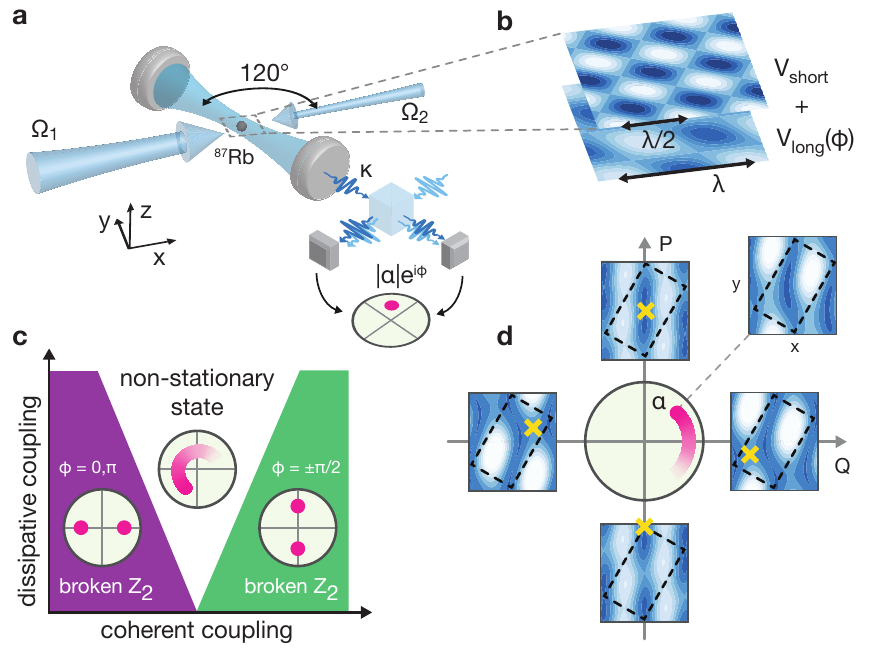}
\caption{\textbf{Non-stationary lattice in a dissipative atom-cavity system}. \textbf{a}, A synthetic crystal is realised by a $^{87}$Rb BEC self-organizing in an optical cavity. Photons scatter in the cavity from a pair of imbalanced transverse beams (with Rabi rates $\Omega_{1,2}$) and leak from the cavity at a rate $\kappa$. A heterodyne setup records in real time the coherent cavity field $\alpha = |\alpha|e^{i\phi}$. \textbf{b}, The interfering light fields create an optical lattice potential that can be decomposed into a component with a short lattice period ($\lambda/2$) and a large period ($\lambda$), with $\lambda$ being the wavelength of light. The spatial position of the long period lattice is determined by the time phase $\phi$ of the scattered field, which is shifted by cavity dissipation. \textbf{c}, Conceptual phase diagram. A finite photon number in the real ($Q$) or imaginary ($P$) quadrature of the cavity field breaks each a $\mathbb{Z}_2$ symmetry, corresponding to different centrosymmetric crystals in real space. In absence of dissipation, increasing the coupling strength results in a structural phase transition between the symmetry broken states. Dissipative coupling mixes the optical quadratures, and a third regime appears in between these states, where the cavity field undergoes persistent oscillations. \textbf{d}, In this regime, the chiral evolution of $\alpha$ drags the atoms through the $\mathbb{Z}_2$ crystal configurations. The dashed rectangle on top of the lattice potentials highlights the unit cell. After a full revolution, the wavefunction's centre of mass (yellow crosses) has been shifted by one unit cell, closing a pump cycle.}
\label{fig1}
\end{figure}

\begin{figure*}[t]
\includegraphics[width = \textwidth]{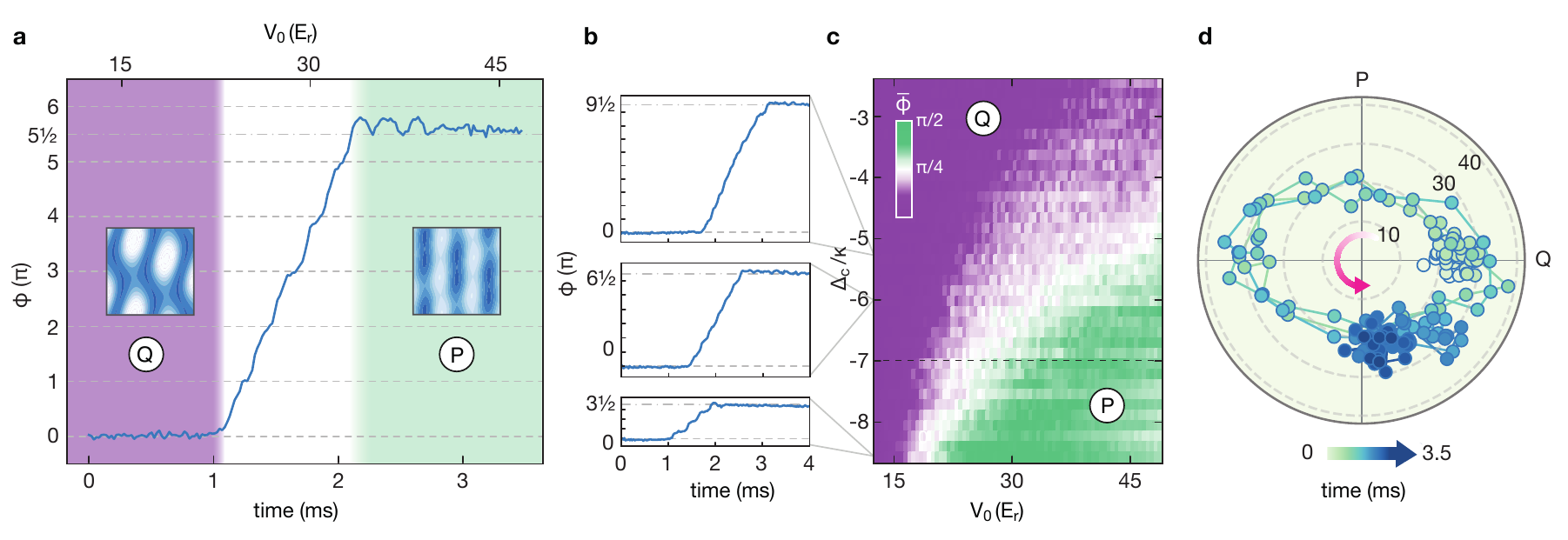}
\caption{ \textbf{Emergent dynamics of the intracavity field.} \textbf{a}, Single trajectory of the time evolution of the intracavity field phase $\phi$, recorded for at $|\Dc/\kappa|=6.94(2)$.  The coupling $V_0$ ramps linearly from 12(1) to 49(5) $E_r$. 
The insets show the calculated real space potentials associated to a purely real or imaginary steady-state field quadrature ($Q$ and $P$, respectively). \textbf{b}, Same as \textbf{a}, but for different dissipative coupling strengths. From top to bottom $|\Dc/\kappa|=5.31(2),5.99(2),8.50(2)$. \textbf{c}, Phase diagram. Each horizontal line shows the mean phase $\bar{\phi}$ obtained from averaging over 50 experimental runs. The polar angle is folded into the first quadrant, $\bar{\phi}\in[0,\pi/2[$, such that an increasing phase appears as $\pi/4$ on average. Horizontal dashed line indicates the value of $\Delta_c/\kappa$ at which the trace in \textbf{a} was recorded. \textbf{d}, Polar plot of the cavity field $\alpha$ in quadrature space, same data as in \textbf{a}. The radii indicate the modulus of $|\alpha|$ and the colormap follows the time axis.}
\label{fig2}
\end{figure*}

The location of the Wannier centre in the unit cell defines the polarisation of the system, which is the real space manifestation of the band topology~\cite{resta2000manifestations,vanderbilt1993electric,resta1994macroscopic}.  
The pumping current can thus be understood as the periodic evolution of the polarisation due to the motion of the Wannier centre~\cite{vanderbilt1993electric}. 
In the case of a centrosymmetric crystal, the Wannier center may either sit on or in-between the inversion center, defining two different $\mathbb{Z}_2$ classes. When evolving continuously between the different centrosymmetric classes without closing the band gap, a current must occur.
This argument can be extended to pumping mechanism in two-dimensional environments ~\cite{resta2007theory}.

In this work, we show that geometrical pumping can emerge self-consistently \textit{without} applying any external time-dependent drive.  We demonstrate this employing a Bose-Einstein condensate coupled to a dissipative cavity, where the dynamics is governed by the self-consistent interplay between matter and light-fields~\cite{mivehvar2021cavity}. 
The geometrical pumping  process is set in motion by dissipation after crossing a dynamical instability point of a self-organized phase, which marks a transition from a steady-state to a self-oscillating state~\cite{jenkins2013self}. 
Coupling internal degrees of freedom to a cavity, a self-oscillation between density and spin waves has recently been observed in a spinor BEC ~\cite{Dogra2019diss}.
Such emerging dynamical phases and persistent oscillations are a distinctive feature of non-Hermitian and non-reciprocal physics~\cite{el2018non, fruchart2021non}. 

In our experimental setup, a Bose-Einstein condensate (BEC) of $^{87}$Rb is placed within an optical resonator. We illuminate the atoms with a pair of counter-propagating  laser beams, having different Rabi rates $\Omega_{1,2}$, and that intersect the cavity mode at a 60\textdegree\,angle~(Fig.~\ref{fig1}a). Collective light scattering from the transverse beams into the cavity is accompanied by a spatial symmetry breaking in the BEC~\cite{Baumann2010}, which forms a density wave by self-organising in the optical lattice generated by the inference of the cavity field with the transverse beams. The resulting potential $V_\mathrm{lattice} = V_\mathrm{short} + V_\mathrm{long}(\phi)$ consists of a short-spaced two-dimensional lattice $V_\mathrm{short}$, fixed in space, and a lattice $V_\mathrm{long}$ with doubled periodicity, whose position depends on the time phase $\phi$ between the transverse beam and the cavity field~(Fig.~\ref{fig1}b). 
The total lattice has two $\mathbb{Z}_2$ symmetry classes, which are set by $\phi$: for $0$ or $\pi$  the scattered field is in-phase with the coupling beams and for $\pm \pi/2$ it out of phase, i.e. the atoms couple either to the real or the imaginary quadrature of the cavity field, respectively~(Fig.~\ref{fig1}c). 
Hence, the quadratures $Q$ and $P$ of the cavity field directly reflect the unit cell polarisation of the lattice~\cite{li2021first}. Which of the two symmetry classes is realized can be controlled by the strength of the transverse beams.

\begin{figure}[h!]
\includegraphics[width = \columnwidth]{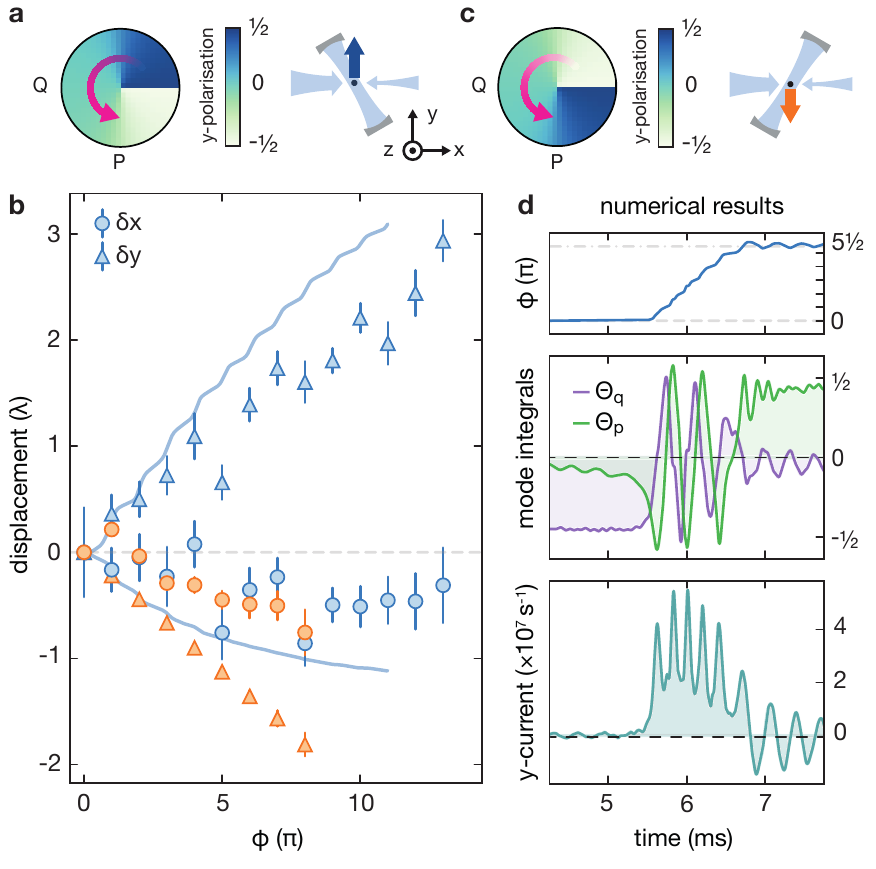}
\caption{\textbf{Self-consistent atomic pump.}  \textbf{a}, Quadrature space and corresponding unit cell polarisation (along the $y$ direction). The singularity is at $|\alpha|=0$. A $2\pi$ rotation of $\alpha$ adds a polarisation quantum to the system, displacing the atoms (right blue arrow). \textbf{b}, Centre of mass displacement as a function of phase winding numbers. Triangles and circles indicate movement along $x$ and $y$, i.e. parallel and orthogonal to the transverse beams. Solid lines indicate theoretical displacements obtained from a numerical simulation. Orange points refer to a second cavity. For the first cavity $|\Dc/\kappa|=4.08(8)$ and a linear ramp from $V_0=0$ to 82.8 $E_r$, for the second cavity $|\Dc/\kappa|=3.00(2)$ and a $V_0$ ramp from 0 to 11(1) $E_r$. Error bars are standard deviations. \textbf{c}, The second cavity mirrors the potential in real space, inverting the polarisation and thus the direction of the current (orange points in \textbf{b}). \textbf{d}, Results from numerical simulation at $|\Dc/\kappa|=12$ and a linear ramp from 0 to 24 $E_r$ in 10 ms. From top to bottom: the photon phase $\phi$, expectation values of the overlap integrals $\Theta_q$ and $\Theta_p$, particle current along $y$.}
\label{fig3}
\end{figure}

Yet, in the open system the steady state response of the cavity field results in a phase shift $\phi_\kappa=\tan^{-1}(\Dc/\kappa)$, that depends on the ratio between the detuning of the laser frequency from cavity resonance, $\Dc$, and the photon dissipation rate $\kappa$, as expected for an oscillator driven close to resonance. This phase shift leads to a local distortion of the effective potential minima and the equilibrium position of the atoms is shifted with respect to the situation with $\Dc\gg\kappa$. This distortion reflects the dissipative coupling between the cavity quadratures. 
If the system is prepared in vicinity to the transition point between the two  $\mathbb{Z}_2$ symmetry classes, the corresponding eigenfrequencies become comparable. For sufficiently strong dissipative coupling these modes hybridize at an exceptional point, encompassing one mode that experiences gain and thus leads to a dynamical instability~\cite{Dogra2019diss,chiacchio2019dissipation,buvca2019dissipation}. In our case, the system is dragged in a chiral motion in quadrature space around a singularity and simultaneously through the $\mathbb{Z}_2$ classes of the crystals.

We detect the emergence of this oscillation in the real-time recording of the quadratures of the cavity field.
The cavity detuning $\Dc$ and the power of the transverse beams, measured in terms of the standing wave potential depth $V_0$, are used as experimental control parameters.  
In our experimental protocol we increase the power of the transverse beams linearly in time while keeping the detuning $\Dc$ (i.e. $\delta_\kappa$) at a fixed value. Fig.~\ref{fig2}a shows a trace of a heterodyne measurement of the cavity field phase $\phi$.  
At the beginning of the ramp, the system is strongly coupled to the real quadrature $Q$, while the imaginary quadrature $P$ is prevailing at the end. In these limiting cases, the phase takes on a defined steady-state value that is determined by the dominant coupling, indicating an initially broken $\mathbb{Z}_2$ symmetry.
Inbetween these steady states, the dynamical instability prevails, and we observe a monotonic increase of the phase $\phi$ in a step-like fashion. This region of self-oscillation widens and the number of windings increases as we repeat the experiment at progressively smaller $|\Dc|$, (Fig.~\ref{fig2}). This is consistent with dissipative effects becoming more dominant as  $\Dc/\kappa\rightarrow0$. In Fig.~\ref{fig2}c we have merged the data obtained for a wide range of detunings $\Dc$, to a phase diagram. The heterodyne data plotted in optical quadrature space (Fig.~\ref{fig2}d), highlights the chiral dynamics of the intracavity coherent field $\alpha$. The ellipsoidal shape, as much as the step-like behaviour of the time evolution of $\phi$, is a direct consequence of the self-consistency, as we will detail in the following. 

An explicit connection between quadratures and crystal structures comes from a microscopic picture of the optical potentials. The short period lattice is generated by the standing waves
\begin{equation}
V_\mathrm{short} = V_0 \cos^2(\kt\mathbf{r}) + \hbar U_0 \ad\a \cos^2(\kc\mathbf{r}),
\end{equation}
where the momenta $\kt$ and $\kc$ (with $|\k|=k=2\pi/\lambda$) are oriented along the transverse coupling beams and the cavity axis, respectively. The period of the potential is set by the wavelength $\lambda = 780$ nm of the light, which we detune by $\Da=+2\pi\times70.3(1)$ GHz with respect to the D$_2$ transition frequency of $^{87}$Rb. At this detuning, spontaneous scattering is largely suppressed and the potential depths can be approximated by off-resonant two-photon scattering, yielding $V_0=\hbar\Omega_1\Omega_2/\Da$ for the lattice formed by the coupling beams and $\ad\a\hbar U_0=\ad\a \hbar g^2/\Da$ for the intra-cavity lattice. The latter is proportional to the intra-cavity photon number $\ad\a$, with $\a$ ($\ad$) the photon annihilation (creation) operator. Here, $g$ is the single photon vacuum Rabi rate of the cavity~\cite{haroche2006exploring}. The interference lattice is given by
\begin{equation}
V_\mathrm{long}= \hbar \eta_q \hat q\cos(\kt\r)\cos(\kc\r) + \hbar \eta_p \hat p\sin(\kt\r)\cos(\kc\r),
\label{eq:Vr}
\end{equation}
where $\eta_{q,p} = g(\Omega_1 \pm\Omega_2)/\Da$ are the two-photon scattering rates between the transverse beams and the cavity mode. The operators of the quadratures of the cavity field are given by $\hat q = (\a+\ad)/2$ and $\hat p= (\a-\ad)/2i$. A finite cavity field ($\braket{\a}\neq0$) is the signature of a self-organised phase ~\cite{Baumann2010}, in which the atomic density is periodically modulated in space, such that the Bragg condition for the scattered light is fulfilled. This density wave breaks the discrete $\mathbb{Z}_2$ symmetry of the Hamiltonian~\cite{baumann2011exploring}, which is invariant under the operation $(\a,x) \rightarrow (\a e^{i\pi},x+\lambda)$. The phase transition happens above a critical coupling strength of $\Omega_{1,2}$.  

From Eq. \eqref{eq:Vr}, it is evident that $\braket{\hat q} =Q=0$ or $ \braket{\hat p} = P=0$ result in different lattice structures.
Couplings to both optical quadratures is facilitated by imbalanced Rabi rates ($\Omega_1\neq\Omega_2 $)~\cite{li2021first,fan2020atomic}. The resulting asymmetry between $\eta_{q,p}$ explains the different cavity field amplitude in the two crystalline phases, leading to the ellipsoidal trajectory in Fig.~\ref{fig2}d. 
For the chosen atomic detuning $\Da>0$, the self-organised phase corresponding to the real quadrature $\braket{\hat q}$ vanishes for increasing transverse lattice depth~\cite{zupancic2019}. Therefore, increasing the coupling strength results in lowering $\braket{\hat q}$ and leads to a competition with the crystal structure with $\braket{\hat p}\neq0$ .

Photons leaking out of the cavity provide real-time access to the atomic distribution, probing the phase transitions~\cite{mekhov2007probing} and the atomic currents~\cite{laflamme2017continuous}. In our experiment, the cavity field phase is given by (see Methods)
\begin{equation}
\phi= \phi_\kappa + \phi_{qp} = \tan^{-1} \biggl(\frac{\Dct \eta_p \To - \kappa \eta_q \Te }{ \kappa \eta_p \To + \Dct \eta_q\Te}\biggr),
\label{eq:phi}
\end{equation}
containing the information on the atomic distribution via the expectation values of the overlap integrals $\Te = \braket{\cos(\kt\r)\cos(\kc\r)}$ and $\To = \braket{\sin(\kt\r)\cos(\kc\r)}$. Here, the bare cavity detuning $\Dc$ has been replaced by $\Dct$, which is the dispersively shifted resonant frequency of the cavity (see Methods).
Therefore, the data in Fig.~\ref{fig2} provide an indirect measurement of the atomic displacement via the overlap integrals in Eq.\eqref{eq:phi}, i.e. a change in the phase of the light field signals that the position of the atomic cloud has shifted in real space.

The chiral evolution of $\alpha$ (e.g. Fig.~\ref{fig2}d) demonstrates the presence of a pumping mechanism, since it implies a monotonous displacement of the atoms in a non-vanishing optical potential. As further confirmation, we perform \textit{in-situ} measurements of the  center-of-mass position of the atomic cloud. We observe that the cloud moves in real space, mainly in the direction orthogonal to the transverse beams, with a displacement that correlates with the observed windings of the photon phase $\phi$ (Fig.~\ref{fig3}).
We observe a displacement of less than one lattice site per pump cycle. We attribute this to the inhomogeneous filling of the lowest band~\cite{lu2016geometrical} and to reduced adiabaticity for increasing pump speeds. However, the movement fully correlates with the number of windings, demonstrating robustness of the transport (see Methods).

The angle between the transverse beams and the cavity is crucial for pumping. 
For deep $V_0$, the atomic distribution resembles a collection of 1D bipartite chains along the $y$ direction, which reminds of the one-dimensional Rice-Mele pump, although the measured lattice depths do not fully justify a tight binding approximation. In this analogy, the angle between the beams breaks the inversion symmetry of the unit cell, contrary to the case of orthogonal interference. 

The direction of the atomic current can in principle be inverted by winding around the singularity with opposite chirality in parameter space.
In our system, the chiral evolution of $\phi$ is fixed by the intrinsic non-Hermitian origin of the dynamics. We can however change the sign of the singularity by inverting the system in real space. We use a second cavity which is mounted realising a $y\rightarrow -y$ transformation, that flips the unit cell polarisation (Fig.~\ref{fig3}c). By coupling the atoms to this cavity, we indeed observe an inversion of the pumping direction (Fig.~\ref{fig3}b).

We have furthermore performed numerical simulations of the open system Hamiltonian. We integrate the self-consistent equations for the light field and the BEC (see Methods), which also gives access to experimentally inaccessible observables such as the overlap integrals of Eq.\eqref{eq:phi} (Fig.~\ref{fig3}d). The numerical results qualitatively match the experimental data, confirming both the field dynamics and the atomic current associated with every phase step (theory lines in Fig.~\ref{fig3}b).
The global nature of the coupling between atoms and light fields allows us to capture the essential mechanism in a mean field description, and the instability is fully captured adding photon dissipation to the time-independent Hamiltonian. In the theoretical approximation of an infinite homogeneous system, the non-stationary state will pump for infinitely long times.

\begin{figure}[h!]
\includegraphics[width = \columnwidth]{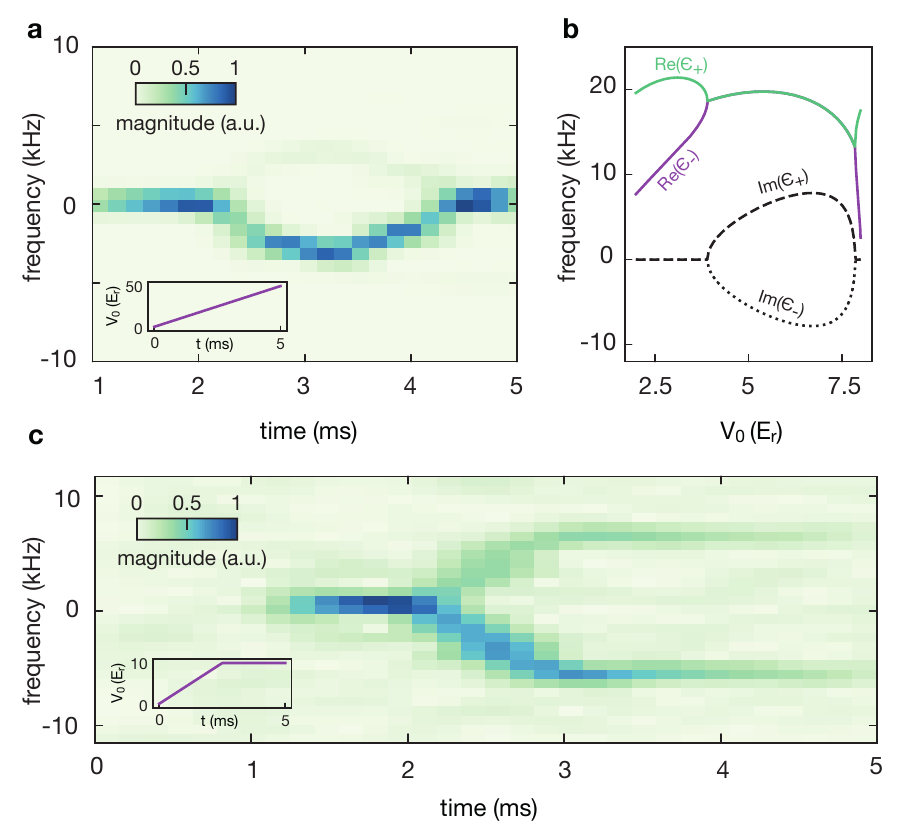}
\caption{\textbf{Non-Hermitian spectra.}  \textbf{a}, Spectrogram of the heterodyne data. A dominant red sideband signals the pumping regime. Data taken at $|\Dc/\kappa|=5.10(2)$ with a linear lattice ramp from $V_0=9.8(9)$ to 49(4) $E_r$. 
\textbf{b}, Eigenvalues $\epsilon_\pm$ of the two mode non-Hermitian toy model. The dissipation leads to the coalescence of the real part and the appearance of imaginary frequencies in the spectrum. $|\Dc/\kappa|$ is the same of \textbf{a}.
\textbf{c} We also measure the frequency at a fixed coupling, by ramping the system in the dynamical instable region and holding the parameters. In this case, the frequency stays constant as long as there is a detectable cavity field.} 
\label{fig4}
\end{figure}

A complementary observable to the chiral evolution in quadrature space is given by the detection of distinct sidebands in the heterodyne spectrum, i.e. the scattered photons have a frequency different from the ones provided by the transverse beams (Fig.~\ref{fig4}a). The data show how the atoms are accelerated and decelerated when the system enters and exits the non-stationary phase, respectively.
Considering the self-organised crystal as a lattice satisfying Bragg's law, this difference can be understood as the Doppler shift associated with the Bragg planes moving as a whole. Energetically, a dominant red sideband is a signature of kinetic energy transferred to the atoms.
We can get a qualitative insight of this frequency shift by expanding the atomic wave function in momentum space. We consider a toy model consisting of only two modes (see Methods), which already captures the oscillation feature of the integrals shown in Fig.~\ref{fig3}d. In this picture, the dissipation is an off-diagonal coupling term in the dynamical matrix of the system, that leads to a coalescence of its eigenmodes (Fig.~\ref{fig4}b).
The appearance of an imaginary part in the spectrum is a feature of optical systems with gain and loss channels and, in general, of non-Hermitian models showing exceptional points~\cite{el2019dawn,ashida2020non}. Our system is unique in mapping such emerging features from the photonic space to a pumping potential.

In a different experimental protocol, we also measure the emerging frequency at fixed coherent coupling strength. To this aim, we first ramp up the lattice strength $V_0$, and subsequently keep $V_0$ fixed during a hold time (Fig.~\ref{fig4}c). 
We observe that the frequency increases while $V_0$ is driving the system into the dynamical region and then stays at a constant value, indicating that the system indeed undergoes persistent oscillations. The signal eventually vanishes due to atom losses. The emerging motion, together with the self-organised lattice, breaks simultaneously time and space translation symmetry. The former can be linked to the concept of continuos dissipative time crystals~\cite{buvca2019dissipation,booker2020non,Dogra2019diss,kongkhambut2022observation}, in contrast to discrete realisations in systems with a time-dependent parametric drive~\cite{cosme2019time,kessler2021observation}. However, the mechanism presented in this work features a topological 
gap. We therefore expect the superfluid system to be protected from energy absorption in the oscillating steady state. 

The observation of a self-consistent pump rises interesting questions that call for future investigations. 
Besides the concept of a topological time crystal, a natural extension would be a generalisation to insulating systems. Coupling the cavity field to a self-organised Mott insulator~\cite{landig2016quantum} or Fermi gas~\cite{zhang2021observation}, the transported charge would be quantised and the cavity phase would directly reflect the topological index~\cite{mivehvar2017superradiant}.
In addition, applying our scheme to a spinor quantum gas~\cite{landini2018formation,kroeze2018spinor,morales2019two} may generate self-driven $\mathbb{Z}_2$ spin pumps.

\FloatBarrier

\section*{Acknowledgments}
We are grateful to Alexander Frank for his support with the heterodyne electronics and we thank Oded Zilberberg and Nicola Spaldin for discussion. 
We acknowledge funding from SNF: project numbers IZBRZ2\_186312, 182650 and 175329 (NAQUAS QuantERA) and NCCR QSIT, from EU Horizon2020: ERCadvanced grant TransQ (project Number 742579) and ITN grant ColOpt (project number 721465).

\section*{Author contributions}
D.D, A.B. and X.L. prepared the experiment, D.D., A.B., X.L. and S.H. took and analysed the data. D.D. performed the numerical simulations. T.D. and T.E. supervised the work.
All authors contributed to discussions of the manuscript.

\section*{Author information}
The authors declare no competing financial interests. Correspondence and requests for materials should be addressed to T.E. (esslinger@phys.ethz.ch).

\newpage

\section*{Methods}\label{SI}

\renewcommand{\figurename}{Extended Data Figure}
\setcounter{figure}{0}

\subsection*{Preparation of the Bose Einstein Condensate (BEC)}

We prepare a gas of $^{87}$Rb in an optical dipole trap.  We reach degeneracy of the bosonic gas via optical evaporation, preparing an almost pure BEC of $N=2.5(4)\times 10^5$ $^{87}$Rb atoms in the hyperfine state $\ket {F=1, m_F=-1}$. The final crossed optical dipole trap consists of two laser beams at a far-detuned wavelength of $1064$~nm  that create an ellipsoid-shaped trap with the three trap frequencies $[\omega_x,\omega_y,\omega_z] = 2\pi\times[89(3),74(1),224(2)] \text{Hz}$. These trap frequencies  set a time frame of around $10$~ms in which we can observe displacements of the cloud without relaxation due to the harmonic confinement.

A magnetic offset field of $\sim25$~G is applied to avoid spin-dependent effects on self-organization as e.g. superradiance to the orthogonally polarized cavity mode that is significantly detuned due to the birefringence of our cavity as we studied in~\cite{morales2019two}.

\subsection*{Imbalanced Transverse Beams}

To engineer the couplings $\eta_{q, p}$ to the two different quadratures of the light field we use imbalanced transverse beams as  explained and utilized in~\cite{li2021first}. The BEC is placed at a distance of around 4 mm from an in-vacuo retro-reflecting mirror. Tuning the imbalance is achieved by mechanically shifting the position of the focussing lens along the optical axis of the pump beam. We extract the imbalance from the experimental data by comparing the threshold of the two distinct self-organized phases (coupling either to the field quadrature $P$ or $Q$) with numerical calculations~\cite{li2021first}.  
The imbalance is quantified by the imbalance parameter
\begin{equation}
\gamma=\sqrt{\frac{\Omega_1}{\Omega_2}},
\end{equation}
with $\Omega_{1,2}$ being the Rabi frequencies of the two plane waves counterpropagating along $x$. The lattice depth $V_0$ is calibrated using Raman-Nath diffraction at the standing-wave component (that couples to the $P$ quadrature of the cavity field) together with the extracted imbalance. For the data in Fig. 2,3 we use an imbalance parameter of $\gamma=1.36(5)$.

\subsection*{Heterodyne measurement}

Balanced optical heterodyne detection is used to record  the amplitude of the intracavity field and its phase with respect to the transverse beam lattice. A local oscillator (LO) laser field is combined on a beam splitter with the light field leaking from one cavity mirror and guided together onto the two photodiodes of a balanced photodetector. The signal is afterwards mixed down to a moderate radio frequency of $300\,\text{kHz}$ with a home build IQ filter device. The LO is generated by shifting the frequency of the same laser source as the transverse beam with an acousto-optical modulator by $50\,\text{MHz}$. To compensate for phase drifts introduced by sending the light fields through separate optical fibers, we lock their relative phase after passing the fibers. The $300\,\text{kHz}$ signal data is recorded with a computer interfaced oscilloscope (PicoScope 5444b). The signal is then processed by a software which extracts the quadratures via a digital IQ mixer and applies a low-pass filter using a binning window of $1\times10^{-4}$ s. All related radio frequency signal sources are phase locked to a $10\,\text{MHz}$ GPS frequency standard which has a fractional stability better than $10^{-12}$ to minimize the technical phase noise in the heterodyne detection system. More technical details and design considerations are described in detail in~\cite{li2021first}.

The measured absolute value of the phase $\phi$ of the intracavity field is dominated by technical fluctuations in-between experimental repetitions like e.g. fiber length drifts, air density changes and unsychronized start of experimental sequences. Nevertheless, the value relative to the pump field is fixed: When the system enters the superradiant phases coupled to the quadrature $Q$ ($P$), the phase $\phi$ is locked to either 0 or $\pi$ ($\pi/2$ or $-\pi/2)$. However, since the lines in the phase diagrams consist of independent measurements, we plot $\phi$ modulo $\pi/2$ in Extended Data Fig.~\ref{SIfig1}. In the main text we focus on the dynamics between the quadratures $P$ and $Q$, where we set the starting point for the physical processes to the positive $Q$ quadrature.  

To calibrate the amplitude of the intracavity light field, we carry out Raman-Nath diffraction analogous to the calibration of the transverse beam by applying to the cavity a short coherent on-axis probe.

\subsection*{Cavities and Detunings}

We can reverse the displacement of the atomic cloud by performing the experiment with two physically distinct cavities that set different directions for the cavity wave vector (see Fig. \ref{fig3}) and also show different rates of dissipation. Both cavities are near-planar with a vacuum Rabi coupling of $[g_1, g_2]=2\pi\times[1.95,1.77]\, \text{MHz}$ and a mode waist of $[48.7,50.4]\,\mu\text{m}$ which is much larger than the atomic cloud itself. This also justifies to model the system with a transversally constant cavity lattice field within the atomic cloud.  Cavity 1 with dissipation rate $\kappa_1 =2\pi\times 147(4)\text{kHz}$ is used for most experiments demonstrated in the main text (see Fig. \ref{fig2} - \ref{fig4}). Cavity 2 is used for the orange data points in Fig. \ref{fig3} and has a higher dissipation rate of $\kappa_2 = 2\pi\times800(11)\,\text{kHz}$.
 
 All measurements are done far in the dispersive regime where the cavity resonance is detuned by $\Delta_a=+2\pi\times 70\,\text{GHz}$  with respect to the Rubidium $\text{D}_2$-line. The detuning between the frequency of the transverse beams and the cavity is tuneable within tens of MHz via an electro-optical modulator. The length of the cavity is stabilised using a low amplitude, far-detuned additional laser field at $830 \, \text{nm}$,  allowing a constant feedback on the cavity length while having a negligible effect on the atomic cloud.

\subsection*{Imaging of the atomic cloud}

Besides the detection of the light field leaking out of the cavities we use resonant absorption imaging to measure the atom number, temperature, momentum distribution, and centre-of-mass position of the atomic cloud.  The physical imaging is done through the viewport below the vacuum chamber by an imaging objective built of two spherical lenses with an effective numerical aperture of $\text{NA}=0.15$. This limits the resolution at our imaging wavelength of $\lambda_{\text{D}_2}=780 \, \text{nm}$ to $\Delta X_\text{res}=2.5 \, \mu\text{m}$. The final image of the atoms is recorded on a CCD camera (Pointgrey Grashopper CCD)  with a pixel size of $4.50 \, \mu\text{m}$ and $1452\times1932$ pixels. This results together with the magnification of the imaging system in an effective pixel size of  $2.25 \, \mu\text{m}$, which allows for precise centre-of-mass detection \textit{in situ}, while still allowing for time-of-flight measurements with the same imaging system.

Temperature, number of atoms, and the momentum state populations are extracted from absorption images taken after $25 \, \text{ms}$ of ballistic expansion of the atomic cloud. The displacement measurements of the cloud's centre-of-mass position shown in Fig. \ref{fig3} are instead extracted by fitting a 2D-Gaussian to \textit{in-situ}  absorption images taken within the dipole trap. Shot-to-shot fluctuations of the cloud position (standard error of $4.54\,\mu\text{m}$) are dominated by position fluctuations of the dipole trap. However, by independently imaging the position of the dipole beam and correlating  the trap position with the cloud position, we reduce the error by a factor of $\approx\times4$. This procedure, together with the large number of measurements, results in the error bars  displayed in Fig. \ref{fig3}. 

\subsection*{Measurement protocol and data selection}

The data shown is taken by setting the cavity detuning $\Delta_c$ to a fixed value and ramping the transverse beam within $5\,\text{ms}$ linearly from zero to a desired transverse beam lattice depth $V_0$. 

\begin{figure}[h!]
\centering
\includegraphics[width= \columnwidth]{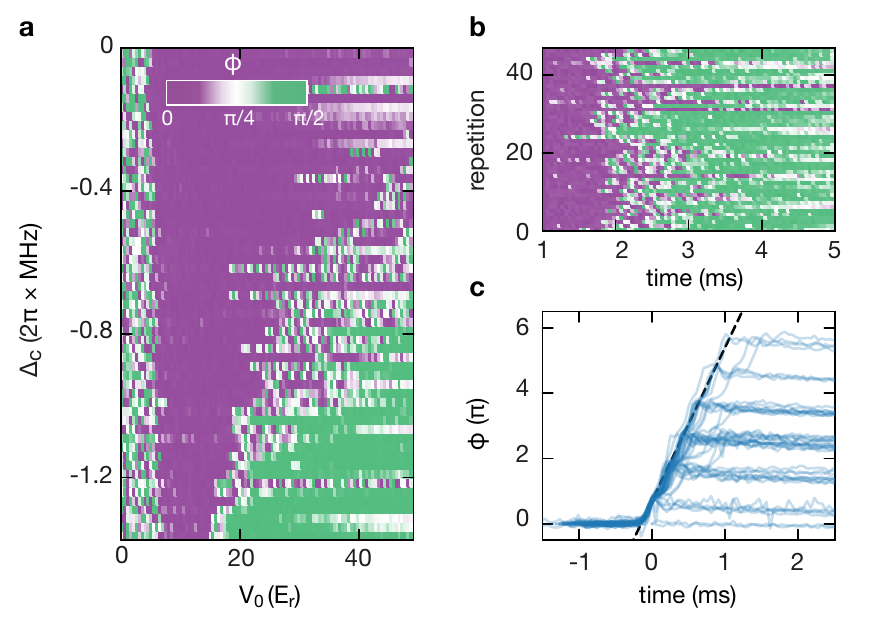}
\caption{\textbf{Non-averaged phase diagram and repeated measurement results.} \textbf{a}, Phase diagram using the phase data $\phi(t)$ from the heterodyne detector by varying cavity detuning $\Delta_C$ of cavity 1. The two different self-organised phases can be well observed for values around $\phi = 0$ and $\phi = \pi/2$. At low transverse beam fields the system shows no self-organisation and $\phi$ is not well defined. Between the two self-organised phases the dynamical phase with varying $\phi(t)$ is visible. \textbf{b}, Many repetitions of the same trace with constant $\Delta_C=-1.1 \, \text {MHz}$. The extent of the instability region varies slightly at each repetition.\textbf{c}, Same data as in \textbf{b}, but shifted in time such that the onset of pumping coincides for all traces. The dashed line as guide to the eye illustrates that the rate at which the phase evolves is robust.}
\label{SIfig1}
\end{figure}

While Fig.~\ref{fig2} shows a phase diagram resulting from averaging 50 experimental repetitions, in Extended Data Fig.~\ref{SIfig1}a we display a phase diagram constructed from a single experimental run at each $\Delta_c$. Extended Data Fig.~\ref{SIfig1}b shows the consecutive repetitions of such experimental runs at a fixed detuning $\Delta_c$, from which the averaged phase diagram can be derived. We attribute the observed variations between the different runs to fluctuations in the atom number and in the relative alignment between the cavity mode and the transverse beams.

The displacement data in Fig.~\ref{fig3} shows the correlations between the number of phase windings and the displacement of the cloud. Since the duration of a single phase winding is short ($<0.1\,\text{ms})$ compared to the whole ramp time, the fluctuations shown in Extended Data Fig.~\ref{SIfig1} make it impossible to deterministically choose a number of phase windings. Instead, we choose a final transverse beam lattice depth such that the ramp ends within the dynamic phase, where we perform the \emph{in-situ} absorption imaging to determine the position of the cloud. After recording a large number of repetitions we select \textit{a posteriori} the desired number of phase windings and plot the according displacements in Fig.\ref{fig3}b. One data point displayed for cavity 1 (orange data) contains on average 394 experimental repetitions (ranging between 15 and 1008, depending on the number of phase windings), while one data point displayed for cavity 2 (blue data points) contains on average 34 experimental repetitions (ranging between 8 and 60, depending on the number of phase windings).

\subsection*{Many-body Hamiltonian}
In the following we derive a model that describes the self-organisation within the quadratures $P$ and $Q$ and shows how by including the dissipation channel $\kappa$ the non-Hermitian dynamics occurs. 
With the two counterpropagating plane waves (with the strength $\Omega_1$ and $\Omega_2$) of the transverse beam we define all effective two-photon Rabi frequencies in the dispersive coupling regime and rotating-wave approximation: The transverse beam lattice  $V_0 = \hbar\frac{\Omega_1\Omega_2}{\Da}$, the dispersive shift per cavity photon $U_0 = \frac{g^2}{\Da}$, the coupling to the quadrature $Q$, $\eta_q = \frac{g(\Omega_1+\Omega_2)}{\Da}$ and the coupling to the quadrature $P$, $\eta_p = \frac{g(\Omega_1-\Omega_2)}{\Da}$.
The Hamiltonian in the standard form of~\citeM{maschler2008ultracold} is:
\begin{equation}
\begin{split}
 \hat{\mathcal{H}} &= - \hbar\Dc \ad\a  +  \hat{H}_0 +   \hbar U_0  \ad\a \hat{B} \\
 & + \hbar\eta_q \frac{(\a+\ad)}{2} \hatTe -i  \hbar\eta_p  \frac{(\a-\ad)}{2} \hatTo
  \label{eq:Hmanybody}
 \end{split}
\end{equation}
where $\ad$ ($\a$) are the creation (destruction) operators for the cavity photon field and having defined:

\begin{align}
\label{eq:SIintegrals0}
\hat{H}_0 &=\int \dr \Pt(\r)\biggl( \frac{\mathbf{p}^2}{2 m} + V_\mathrm{trap}(\r) \nonumber  \\
& + \frac{\tilde{g}}{2}|\P(\r)|^2 + V_0  \cos(\kt\r)^2\biggr)\P(\r)   \\ 
\hat{B} & =  \int \dr \Pt(\r) \cos(\kc\r)^2\P(\r) \\
\label{eq:SIintegrals}
\hatTe &  = \int \dr \Pt(\r)  \cos(\kt\r)\cos(\kc\r)\P(\r) \\
\label{eq:SIintegrals2}
\hatTo  & = \int \dr \Pt(\r) \sin(\kt\r)\cos(\kc\r)\P(\r)  
\end{align}
$\kt$ and $\kc$ are the pump and cavity wave vectors respectively (with $|\k|=k=2\pi/\lambda$). The direction of the transverse beam is chosen such that the field with strength $\Omega_1$ is along the x-axis and the field with strength $\Omega_2$ is counterpropagating. The atomic cloud is small enough to neglect the transverse spatial component of the electromagnetic modes. The physical interpretation of the integrals is: $ \hat{H}_0 $ is describing the BEC with collisional interaction $\tilde{g}$ and trapped in the harmonic potential $V_\mathrm{trap}$ and in the optical lattice of the transverse beam, $\hat{B} $ is the overlap with the cavity mode, and $\hatTe$ and $\hatTo$ the overlaps with the two patterns coupling to the different quadratures. These integrals are our choice of order parameter in a phenomenological theory of the self-organisation phase transition.
Note that changing $\mathrm{sign}(\Omega_1-\Omega_2)$ from an Hamiltonian point of view corresponds to flipping $x\rightarrow -x$ and the same is valid for changing the direction of $\kc$. 

\subsection*{Heisenberg equation of motion and dissipation}

To observe the dynamics of the photonic part of the Hamiltonian we look at the Heisenberg equation of motion of the field operator of the light, incorporating also the cavity dissipation $\kappa$: 
\begin{equation}
\frac{\partial \braket{\a} }{\partial t} = -\frac{i}{\hbar}\braket{[\a, \hat{\mathcal{H}}]}-\kappa  \braket{\a},
\end{equation}
where we omitted quantum fluctuations since they equal zero at the mean field level for $\alpha = \braket{\a}$.
We assume a quasi-stationary light field $\partial_t \alpha  = 0$ since the time scales of the light field separate from the motional time scales of the atomic wavefunction. With this we get for the light field:
\begin{align}
\alpha &= e^{i\phi} \sqrt{ \frac{ \eta_q^2 \Te^2 + \eta_p^2\To^2 }{\Dct^2+\kappa^2}}  \nonumber  \\
 &=e^{i\phi_\kappa + i\phi_{qp}} \sqrt{ \frac{ \eta_q^2 \Te^2 + \eta_p^2\To^2 }{\Dct^2+\kappa^2}} .
\label{eq:alpha}
\end{align}

\begin{figure}[ht!]
\centering
\includegraphics[width= \columnwidth]{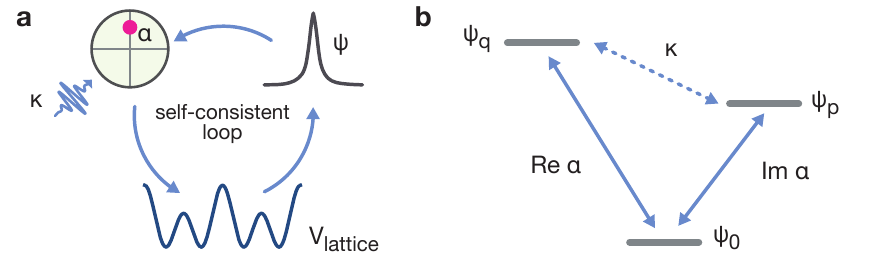}
\caption{\textbf{Schematic representation of the theoretical models.} \textbf{a}, Self-consistent loop between cavity field $\alpha$, optical lattice $V_\mathrm{lattice}$ and wave function $\psi$ as described by the set of equations Eq.\eqref{eq:Hmanybody} and Eq.\eqref{eq:alpha}.\textbf{b}, Minimal model given by the three level momentum expansion of Eq.\eqref{eq:3modewavefunction}. The coherent coupling (solid arrows) mixes the condensate mode $\psi_0$ with the spatially modulated $\psi_{p,q}$, which are then mutually coupled by dissipation (dashed arrows).}
\label{SIfig3}
\end{figure}

The complex light phase $\phi$ is defined in equation \eqref{eq:phi} of the main text and can be separated into $\phi_\kappa=\tan^{-1}(\frac{\Dct}{\kappa})$ and $\phi_{qp}=\tan^{-1}(-\frac{\eta_q \Te}{\eta_p \To})$.
For simpler notation we define $\Theta_i = \braket{\hat \Theta_i}$, $B = \braket{\hat B}$ and $\Dct = \Dc -  U_0 B$, which is the dynamical dispersive cavity detuning due to the changing overlap of the atomic wavefunction with the cavity field. Equations \eqref{eq:alpha} and \eqref{eq:Hmanybody} thus form a self-consistent loop governing the dynamics of the matter and the light fields (Extended Data Fig.~\ref{SIfig3}a).

\subsection*{Equation of motion and low-energy momentum expansion}

To get further intuition for the dynamics of the photon field and the atomic wavefunction, we can construct an atomic field operator from eigenstates $\psi_i$ of the momentum operator:
\begin{equation}
\P = \sum_i \c_i \psi_i,
\end{equation}
where $\c_i$ annihilates a particle in the state $\psi_i$. After substitution in the many-body Hamiltonian~\eqref{eq:Hmanybody} one gets
\begin{equation}
\begin{split}
 \hat{\mathcal{H}} &= - \hbar\Dc \ad\a  + \sum_{i,j} \biggl(H^{ij}_{0} +   \hbar U_0  \ad\a B^{ij} \\
 & + (\a+\ad)\hbar\er  \Theta_{q}^{ij} -i (\a-\ad) \hbar\ei\Theta_{p}^{ij}\biggr) \cd_i\c_j,
 \end{split}
\end{equation}
where the notation $\mathcal{O}^{ij} = \bra{\psi_i} \hat{\mathcal{O}}\ket{\psi_j}$ represents the expectation values of the operators in Eq.\eqref{eq:SIintegrals0}-\eqref{eq:SIintegrals2}.
The integrals defined in Eq.~\eqref{eq:SIintegrals}-\eqref{eq:SIintegrals2} only connect the BEC to momentum families separated by $\kc$, $\kt$ and linear combinations.
In a low-energy approximation it is then appropriate to use the Ansatz
\begin{equation}
\P = \c_0 \psi_0 + \c_q \psi_q + \c_p \psi_p,
\label{eq:3modewavefunction}
\end{equation}
that only uses the three momentum modes $\psi_q=\frac{2}{\sqrt{A}} \cos(\kt\r) \cos(\kc\r)$, $\psi_p= \frac{2}{\sqrt{A}} \sin(\kt\r) \cos(\kc\r)$, and $\psi_0=\frac{1}{\sqrt{A}}$, where $A$ defines the area of the Wigner Seitz cell. In the definition of Eq.~\eqref{eq:3modewavefunction}, each mode is normalised to unity, $\int \psi_i^{\dag} \psi_i =1$ and the field operator is normalised to the total particle number $N$, $\int \Pt \P = \c_0^{\dag}\c_0 + \c_q^{\dag}\c_q + \c_p^{\dag}\c_p =N$. Under this approximation, and neglecting the harmonic trapping potential the collisional interactions, we arrive at the following mean-field equations of motion for the three level system (Extended Data Fig.~\ref{SIfig3}b).:
\begin{equation}
\begin{cases}
\dot \alpha = i \Dct\alpha  - \kappa\alpha  - i \frac{\eta_q}{4}(c_q^*c_0+c_0^*c_q) + \frac{\eta_p}{4} (c_p^*c_0+c_0^*c_p)   
\\
\dot c_0 = i  \frac{\eta_q}{2} \Re(\alpha) c_q - i \frac{\eta_p}{2}\Im(\alpha)c_p
\\
\dot  c_q = -i(\omega_q+\frac{V_0}{4\hbar}) c_q - i \frac{\eta_q}{2} \Re(\alpha) c_0
\\
\dot  c_p   =-i(\omega_p-\frac{V_0}{4\hbar}) c_p - i \frac{\eta_p}{2} \Im(\alpha) c_0
\end{cases}
\end{equation}

\subsection*{Minimal model and non-Hermitian dynamics}
To demonstrate the instability, it is sufficient to reduce the equations of motion further by keeping the $c_0$ population constant ($c_0\sim\sqrt{N}$). This is an appropriate approximation since the relative depletion of the zero momentum state is small. As a result of this approximation, the number of coupled equations is further reduced. Applying another time derivative to $\dot c_{q,p}$, and using the steady state value for $\alpha$, the differential equations for the populations of $\psi_{q,p}$ can be written as 
\begin{eqnarray}
\left(%
\begin{array}{c}
  \ddot c_q  \\
  \ddot c_p 
\end{array}
\right)
=\left(%
\begin{array}{cc}
  \Omega_q ^2 & -K_q  \\
 K_p  & \Omega_p ^2
\end{array}
\right)
\left(%
\begin{array}{c}
  c_q   \\
  c_p 
\end{array}
\right) .
\label{eq:matrix}
\end{eqnarray}

In the above matrix notation, we have defined
\begin{equation}
\Omega_i^2 = -\big(\omega_i \pm \frac{V_0}{4\hbar}\big)^2 - N  \eta_i^2 \big(\omega_i \pm \frac{V_0}{4\hbar}\big) \frac{\Dct}{\Dct^2+\kappa^2} 
\label{omegaeq}
\end{equation}
\begin{equation}
K_i = N\eta_q\eta_p \big(\omega_i \pm \frac{V_0}{4\hbar}\big)  \frac{\kappa }{\Dct^2+\kappa^2} ,
\end{equation}
with $i\in\{q,p\}$ and $\pm$ is positive for $i=q$ and negative for $i=p$. The eigenenergies $\omega_i$ of the q and p mode can be obtained from the free particle dispersion as $\omega_{q,p}=2\omega_r$, with $\omega_r$ being the recoil frequency. The two-photon coupling strengths $\eta_{q,p}$ are defined in the main text via the Rabi rates of the two transverse beams and can be rewritten as $\eta_{q,p} = \frac{g}{\Da} (\Omega_1\pm\Omega_2)=(\gamma \pm \frac{1}{\gamma})\sqrt{U_0V_0}$ to express the explicit dependences on the imbalance parameter $\gamma$, transverse beam lattice $V_0$ and dispersive shift per cavity photon $U_0$.

Note that in Eq. \eqref{omegaeq} the coupling to the $Q$ quadrature becomes weaker for strong enough transverse beam lattice strengths $V_0$, which makes the energies of the $P$ and $Q$ modes cross eventually.
For more details of the differences of the two couplings see Ref.~\cite{zupancic2019}.

The diagonal elements of the matrix describe the soft mode energies. Above a critical pump strength, these matrix elements give rise to growing populations $c_{q,p}$. The off-diagonal elements describe the coupling rate of the two modes via the dissipation. 

From diagonalising the non-Hermitian matrix in Eq.~\eqref{eq:matrix}, we get an analytical expression of the eigenvalues
\begin{equation}
\epsilon^2_{\pm}=\frac{\Omega^2_q+\Omega^2_p}{2} \pm \frac{1}{2}\sqrt{(\Omega^2_q-\Omega^2_p)^2-4K_qK_p}.
\end{equation}
Real and imaginary part of $\epsilon_{\pm}$ describe growth rate and oscillation frequency of the decoupled eigenvectors $c_\pm$, from which the evolution of $c_{q,p}$ can be derived. The populations start oscillating between each other when $(\Omega^2_q-\Omega^2_p)^2<4K_qK_p$, i.e. the dynamics emerges once the dissipative coupling strength $K_i\propto \kappa$ is larger than the energy gap between the two energies. Note that the term $4K_qK_p\propto \Dct^{-4}$ while $\Omega^4\propto \Dct^{-2}$, such that the dissipation can only overcome the energy gap in the regime of $\kappa/\Dct\approx 1$. 
This can be seen in Fig.\ref{fig4}b, when the real part of the eigenvalues coalesce as they are close enough together.

\subsection*{Numerical simulations}

\begin{figure}[h!]
\centering
\includegraphics[width=\columnwidth]{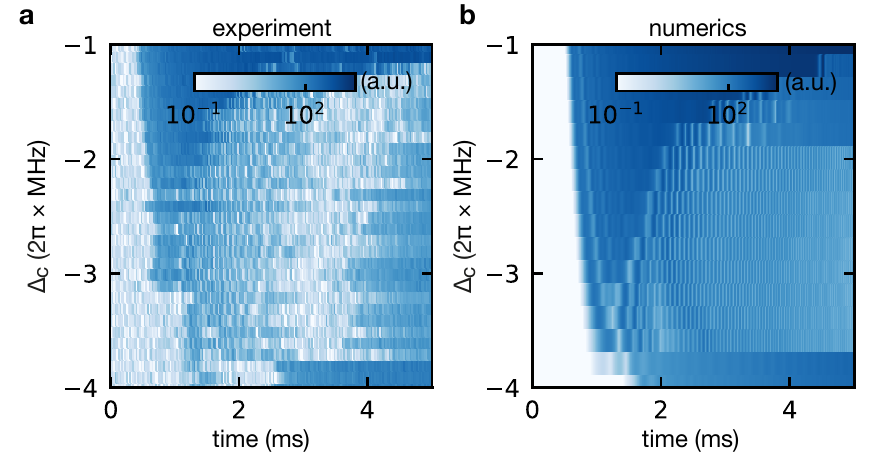}
\caption{\textbf{Comparison of experimental and numerical phase diagrams.} Figures show the amplitude of the intracavity light field dependent on $V_0$ and the cavity detuning $\Dc$ for cavity 2. \textbf{a}, Dataset showing in each row a trace of single experimental realisation for the given cavity detuning $\Delta_c$. The transverse beam lattice is linearly ramped to the final transverse lattice strength $V_0=40\,E_r$ within $5\,\text{ms}$. \textbf{b}, Corresponding simulation of the experimental results with GPE simulation.}
\label{SIfig2}
\end{figure}

Alternatively to the simplified analytical model described above, we numerically integrate the self-consistent many-body Hamiltonian Eq.\eqref{eq:Hmanybody} using a standard symmetrised split-step Fourier method. 
This allows to propagate the equations in time and calculate the dynamics of the system, as well as to obtain the ground state by first performing a Wick rotation. 
The cavity field is assumed in its steady state at each integration time-step, so it can be numerically calculated from the wave function using Eq.\eqref{eq:alpha}. 
All theoretical simulations reproduced in the figures throughout the paper include the trap and inter-particle collisional interactions. 
In the calculations, the transverse extent of the coupling beams and the cavity mode are neglected, such that the initial harmonic confinement is not modified.
We checked the validity of the simulation by comparing the numerical phase diagram to the experimental one (Extended Data Fig.~\ref{SIfig2}), which show qualitative agreement.

The current shown in Fig.~\ref{fig3}d is calculated from the probability current 
\begin{equation}
\mathbf{j}(t) = -i\frac{\hbar}{2m}(\psi^*\mathbf{\nabla}\psi-\psi\mathbf{\nabla}\psi^*)
\end{equation}
integrated over the whole system.

The polarisation of Fig.~\ref{fig3}a,c is calculated by obtaining the ground state wave function in the unit cell at different choices of $\alpha$. Since this calculation was performed in periodic boundary conditions (PBC), the centre of mass has to be extracted using~\citeM{resta1998quantum}. The polarisation
\begin{equation}
P_x = \frac{e}{2\pi L^2} \Im\log\bra{\psi}e^{-i \frac{2\pi}{L} \hat{X}} \ket{\psi}.
\end{equation}
is the phase of the expectation value of $\hat{X}$ (the position operator) in PBC, which, in the words of reference~\citeM{resta2000manifestations} is `just a Berry's phase in disguise'.
Here, $L$ is the linear system size and the charge is set to $e=1$ for neutral atoms. 

\section*{Data availability}
The data to reproduce the figures of this study are available in the data repository of ETH Zurich’s Research Collection (www.research-collection.ethz.ch) under the DOI 10.3929/ethz-b-000547966.

\end{document}